\documentclass[preprint,amsmath,amssymb,aps,showkeys,showpacs]{revtex4}
\usepackage[english]{babel}
\usepackage{graphicx}
\usepackage{graphics}
\usepackage{amsmath}
\usepackage{dcolumn}
\usepackage{amssymb}
\usepackage{bm}

\begin{document}
\title{A central potential with a massive scalar field in a Lorentz symmetry violation environment}
\author{R. L. L. Vit\'oria}
\email{ricardo.vitoria@ufes.br/ricardo.vitoria@pq.cnpq.br}
\affiliation{Departamento de F\'isica e Qu\'imica, Universidade Federal do Esp\'irito Santo, Av. Fernando Ferrari, 514, Goiabeiras, 29060-900, Vit\'oria, ES, Brazil.}

%\author{C. Furtado}
%\email{furtado@fisica.ufpb.br}
%\affiliation{Departamento de F\'isica, Universidade Federal da Para\'iba, Caixa Postal 5008, 58051-900, Jo\~ao Pessoa-PB, Brazil.}

\author{H. Belich}
\email{belichjr@gmail.com}
\affiliation{Departamento de F\'isica e Qu\'imica, Universidade Federal do Esp\'irito Santo, Av. Fernando Ferrari, 514, Goiabeiras, 29060-900, Vit\'oria, ES, Brazil.}

\begin{abstract}
We investigate the behaviour of a massive scalar field under the influence of a Coulomb-type and central linear central potentials inserted in the Klein-Gordon equation by modifying the mass term in the spacetime with Lorentz symmetry violation. We consider the presence of a background constant vector field which characterizes the breaking of the Lorentz symmetry and show that analytical solutions to the Klein-Gordon equation can be achieved.
\end{abstract}


\maketitle

\section{Introduction}

The Standard Model (SM) presents in a unified way the electromagnetic, weak and strong nuclear interactions with the exception of gravitational interaction which makes it an incomplete quantum field theory (QFT). In addition, there are some observations, both from a theoretical and observational point of view, about their predictions. Recent experimental data provided measurements of the proton radius which it is different from the value predicted by the SM \cite{próton}. There is evidence, by observational data, that the fine structure constant, admitted constant by quantum electrodynamics, one of the pillars of the SM, it is changing \cite{cf, cf1}. The SM is also limited in explaining the dark sector of the Universe.

Due to these questions about limitations of the SM, it has arisen in recent years interest in investigating the possibility of a physics beyond the SM. In this context, the Lorentz symmetry violation (LSV) has been extensively explored in QFT since their possible scenarios have provided directions in the search for answers about physical effects of possible underlying physical theories, in which they can not be explained or observed through usual physics. In this sense, based on string theory, Kostelek\'{y} and Samuel dealt with the spontaneous breaking of symmetry through non-scalar fields, in which the vacuum expected value is constant and acquires a tensorial nature which violates the Lorentz symmetry spontaneously \cite{kos}. This possibility of extending the SM was known as the Standard Model Extension (SME) \cite{kos1, kos2}. In recent decades, the LSV has been extensively studied in various branches of physics \cite{vsl, vsl1, vsl2, vsl3, vsl4, vsl5, vsl6, vsl7, vsl8, vsl9, vsl10, vsl11, vsl12, vsl13, vsl14, vsl15, vsl16, vsl17, vsl18, vsl19, vsl20, vsl21, vsl22, vsl23, vsl24, vsl25, vsl26, vsl27, vsl28, vsl29, vsl30, vsl31, vsl32, vsl33, vsl34, vsl35, vsl36, vsl37, vsl38, vsl39, vsl40, vsl41, vsl42, vsl43, vsl44, vsl45, vsl46, vsl47, bel, bel1, bel2, cas, cas1, cas2, cas3, cas4, gaz}. Also, the LSV has been studied in non-relativistic quantum mechanics, for example, Landau-type quantization \cite{bb}, holonomies \cite{bb1}, geometric phase \cite{book} and in relativistic quantum mechanics, in particular, on a scalar field \cite{bb2, bb3, me, me1}.

In this paper, we have investigate the relativistic quantum dynamics of a scalar field subject to a hard-wall potential and to the Coulomb-type and linear central potentials in a spacetime with LSV. Such configuration in the spacetime, which characterizes the LSV, is provided by the direct coupling between the derivative of the field with an arbitrary constant vector field in the Klein-Gordon equation, where we analyze its effects on a scalar field. Thus, we show that it is possible to find out analytically solutions of bound states and to determine the relativistic energy levels for the scalar field in a Lorentz violating background for each case.

The structure of this paper is as follows: in section II, we analyzed the effects of a hard-wall potential on a scalar field in spacetime with LSV caused by the presence of a background constant vector field; in section III, we investigated the effect of a Coulomb-type central potential inserted in the Klein-Gordon equation by modifying the mass term and discuss their effects on a scalar field in an spacetime with LSV; in section IV, we insert a linear central potential in the Klein-Gordon equation by modifying the mass term and determine solutions of bound states for a scalar field in an spacetime with LSV; in section V, by modifying the mass term of the Klein-Gordon equation, we analyzed the effects of a linear plus a Coulomb-type central potential on a scalar field subject to the LSV; in section VI, we present our conclusions.

\section{Effects of the hard-wall potential}\label{SECII}

Effective theories with LSV have been the focus of increasing interest in various physics contexts nowadays. The symmetry that permeates all high energy physics and Lorentz covariance is the basis of the SM of particle physics construction, and so is natural to ask why the interest of this type of violation. Inspired by Refs. \cite{go, cruz}, we can write the equation for a scalar field of the form
\begin{eqnarray}\label{II1}
[\square-g(v^{\mu}\partial_{\mu})^2-m^2]\phi=0,
\end{eqnarray}
where $\square=-\frac{\partial^{2}}{\partial t^{2}}+\nabla^{2}$ is the d'alembertian, $g$ is a coupling constant, $m$ is the rest mass of the scalar field and $v^{\mu}$ is the background vector field responsible by the LSV. It is important to note that the coupling that appears in Eq. (\ref{II1}) it conserves the CPT symmetry, that is, it is a CPT-even coupling \cite{go}. In addition, the LSV non-minimal coupling in the Eq. (\ref{II1}) is analogous to the general structure of the Eq. (6) of the Ref. \cite{kos2}, where the background vector field $v^{\mu}$ is associated with a second-order tensor field $(k_{\varphi\varphi})$ from the Ref. \cite{kos2}: $v^{a}v^{b}\sim-(k_{\varphi\varphi})^{ab}$. Here, we consider the background field vector with the configurations $v^{\mu}=(v^0,0)$ and $v^{\mu}=(0,\vec{v})$. Note that these particular choices do not prevent us from investigating their effects on the scalar field, since they are possible scenarios of LSV from the theoretical point of view.

In this paper, we work in the Minkowski spacetime with cylindrical symmetry ($c=\hbar=1$):
\begin{eqnarray}\label{II2}
ds^2=-dt^2+d\rho^2+\rho^2d\varphi^2+dz^2,
\end{eqnarray}
with $\rho=\sqrt{x^2+y^2}$ being the axial distance.

\subsection{Background vector field with the configuration $v^{\mu}=(v^0,0)$}\label{SUBSECII-1}

From now on, let us make a discussion from the theoretical point of view, where a scalar field is subject to the effects of the LSV given by the presence of the background vector field $v^{\mu}$. In particular, let us consider a background vector field with the following configuration: $v^{\mu}=(a,0,0,0)$, where $a=\text{const.}$. Note that this is a particular scenario of the LSV. In particular case, the Eq. (\ref{II1}) becomes
\begin{eqnarray}\label{II3}
-\frac{\partial^2\phi}{\partial t^2}+\frac{\partial^2\phi}{\partial\rho^2}+\frac{1}{\rho}\frac{\partial\phi}{\partial\rho}+\frac{1}{\rho^2}\frac{\partial\phi}{\partial\varphi}+
\frac{\partial^2\phi}{\partial z^2}-ga^2\frac{\partial^2\phi}{\partial t^2}-m^2\phi=0.
\end{eqnarray}

Let us consider a particular solution to the Eq. (\ref{II3}) given in terms of the eigenvalues of the $z$-component of the angular momentum operator $\hat{L}_{z}=-i\partial_{\varphi}$ and of the eigenvalues of the $z$-component of the linear momentum operator $\hat{p}_{z}=-i\partial_{z}$ as
\begin{eqnarray}\label{II4}
\phi(\rho,\varphi,z,t)=e^{-i\mathcal{E}t}e^{il\varphi}e^{ikz}R(\rho),
\end{eqnarray}
where $l=0,\pm1,\pm2,\pm3,\ldots$, $-\infty<k<\infty$ and $R(\rho)$ is a function of the axial distance. Then, by substituting the Eq. (\ref{II4}) into the Eq. (\ref{II3}), we obtain the ordinary differential equation
\begin{eqnarray}\label{II5}
\frac{d^2R}{d\rho^2}+\frac{1}{\rho}\frac{dR}{d\rho}-\frac{l^2}{\rho^2}R+\alpha^2R=0,
\end{eqnarray}
where
\begin{eqnarray}\label{II6}
\alpha^2=(1+a^2g)\mathcal{E}^2-m^2-k^2.
\end{eqnarray}

The Eq. (\ref{II6}) is the well-known the Bessel differential equation \cite{arf}. The general solution to the Eq. (\ref{II6}) is given in the form: $R(\rho)=C_{1}J_{|l|}(\alpha\rho)+C_{2}N_{|l|}(\alpha\rho)$, where $J_{|l|}(\alpha\rho)$ and $N_{|l|}(\alpha\rho)$ are the Bessel function of first kind and second kind \cite{arf}, respectively. The Bessel function of second kind diverges at the origin, then we must take $C_{2}=0$ in the general solution, since we are interested in a well-behaved solution. Thus, the regular solution to the Eq. (\ref{II6}) at the origin is given by:
\begin{eqnarray}\label{II7}
R(\rho)=C_{1}J_{|l|}(\alpha\rho).
\end{eqnarray}

Let us restrict the motion of the scalar field to a region where a hard-wall potential is present. This kind of confinement is described by the following boundary condition:
\begin{eqnarray}\label{II8}
R(\rho_0)=0,
\end{eqnarray}
which means that the wave function $R(\rho)$ vanishes at a fixed radius $\rho_0$, that is, this boundary condition corresponds to the scalar field subject to a hard-wall potential. The hard-wall potential has been studied in Landau-Aharonov-Casher system \cite{hw}, in a Dirac neutral particle in analogous way to a quantum dot \cite{hw1}, in the relativistic quantum motion of spin-0 particles under the influence of noninertial effects in the cosmic string spacetime \cite{hw2}, in the quantum dynamics of scalar bosons \cite{hw3}, in the Aharonov-Bohm effect for bound states in relativistic
scalar particle systems in a spacetime with a spacelike dislocation \cite{hw4} and in rotating effects on the scalar field in the spacetime with linear topological defects \cite{hw5}. Then, let us consider a particular case where $\alpha\rho_0\gg1$. In this particular case, we can write the Eq. (\ref{II7}) in the form \cite{arf}:
\begin{eqnarray}\label{II9}
J_{|l|}(\alpha\rho_0)\propto\cos\left(\alpha\rho_0-\frac{|l|\pi}{2}-\frac{\pi}{4}\right).
\end{eqnarray}

Hence, by substituting the Eq. (\ref{II9}) into the Eq. (\ref{II7}), we obtain from the boundary condition (\ref{II8}) the relativistic energy levels of the system
\begin{eqnarray}\label{II10}
\mathcal{E}_{k,l,n}\approx\pm\sqrt{\frac{1}{(1+a^2g)}\left[m^2+k^2+\frac{\pi^2}{\rho_0^2}\left(n+\frac{|l|}{2}+\frac{3}{4}\right)^2\right]},
\end{eqnarray}
where $n=0,1,2,3,\ldots$.

We note that the background that characterizes the LSV caused by the presence of the particular vector field influences the dynamics of the scalar field subject to the hard-wall potential through the presence of the parameters associated with the LSV, $g$ and $a$, on relativistic energy levels of the system. We can also note that, by taking $g\rightarrow0$ or $a\rightarrow0$, we obtain the relativistic energy levels in the Minkowski spacetime.

\subsection{Background vector field with the configuration $v^{\mu}=(0,\vec{v})$}\label{SUBSECII-2}

\subsubsection{The axial direction}

Let us consider a vector field which governs the LSV with the following configuration: $v^{\mu}=(0,b,0,0)$, where $b=\text{const.}$. It is important to note that this particular configuration of the vector field that governs the LSV does not arise from the spontaneous breaking of the Lorentz symmetry, since its direction varies. Because it does not have this feature, but still be a type of configuration that breaks the violation of the Lorentz symmetry explicitly, it is treated as an external vector field and not as a background vector field. In this particular case, the Eq. (\ref{II1}) becomes
\begin{eqnarray}\label{II11}
-\frac{\partial^2\phi}{\partial t^2}+\frac{\partial^2\phi}{\partial\rho^2}+\frac{1}{\rho}\frac{\partial\phi}{\partial\rho}+\frac{1}{\rho^2}\frac{\partial\phi}{\partial\varphi}+
\frac{\partial^2\phi}{\partial z^2}-gb^2\frac{\partial^2\phi}{\partial\rho^2}-m^2\phi=0.
\end{eqnarray}
Then, by substituting the Eq. (\ref{II4}) into the Eq. (\ref{II11}), we have
\begin{eqnarray}\label{II12}
(1-b^2g)\frac{d^2R}{d\rho^2}+\frac{1}{\rho}\frac{dR}{d\rho}-\frac{l^2}{\rho^2}R+\beta^2R=0,
\end{eqnarray}
where we define
\begin{eqnarray}\label{II13}
\beta^2=\mathcal{E}^2-m^2-k^2.
\end{eqnarray}
With the purpose of solving the Eq. (\ref{II12}), let us write
\begin{eqnarray}\label{II14}
R(\rho)=\rho^{-\frac{1}{2}\left(\frac{b^2g}{1-b^2g}\right)}f(\rho).
\end{eqnarray}
Then, by substituting the Eq. (\ref{II14}) into the Eq. (\ref{II12}), we obtain the following equation for $f(\rho)$:
\begin{eqnarray}\label{II15}
\frac{d^2f}{d\rho^2}+\frac{1}{\rho}\frac{df}{d\rho}-\frac{\lambda^2}{\rho^2}f+\varepsilon^2f=0,
\end{eqnarray}
where
\begin{eqnarray}\label{II16}
\lambda^2=\frac{1}{4(1-b^2g)^2}(4l^2-4l^2b^2g+b^4g^2); \ \ \ \varepsilon^2=\frac{\beta^2}{(1-b^2g)}.
\end{eqnarray}

Note that the Eq. (\ref{II15}) is analogous to the Eq. (\ref{II5}), then, by following the same steps from the Eq. (\ref{II7}) to the Eq. (\ref{II10}), we have
\begin{eqnarray}\label{II17}
\mathcal{E}_{k,l,n}\approx\pm\sqrt{m^2+k^2+\frac{(1-b^2g)\pi^2}{\rho_0^2}\left[n+\frac{1}{4(1-b^2g)}\sqrt{4l^2-4l^2b^2g+b^4g^2}+\frac{3}{4}\right]^2}.
\end{eqnarray}

The Eq. (\ref{II17}) gives us the energy spectrum of a scalar field subject to a hard-wall potential in an spacetime with LSV in the axial direction. We can note that the effects of the LSV influence the quantum dynamics of the scalar field through the presence of parameters $b$ and $g$. In addition, by making $g\rightarrow0$ or $b\rightarrow0$, we obtain the energy spectrum of a scalar field subject to a hard-wall potential in the Minkowski spacetime.

\subsubsection{z-direction}

From now on, let us consider the background vector field $v^{\mu}=(0,0,0,c)$, where $c$ is a constant. In this particular case, the Eq. (\ref{II1}) becomes
\begin{eqnarray}\label{II18}
-\frac{\partial^2\phi}{\partial t^2}+\frac{\partial^2\phi}{\partial\rho^2}+\frac{1}{\rho}\frac{\partial\phi}{\partial\rho}+\frac{1}{\rho^2}\frac{\partial\phi}{\partial\varphi}+
\frac{\partial^2\phi}{\partial z^2}-gc^2\frac{\partial^2\phi}{\partial z^2}-m^2\phi=0.
\end{eqnarray}

We can follow the steps from the Eqs. (\ref{II3}) to the (\ref{II5}), where we obtain the differential equation
\begin{eqnarray}\label{II18a}
\frac{d^2R}{d\rho^2}+\frac{1}{\rho}\frac{dR}{d\rho}-\frac{l^2}{\rho^2}R+\epsilon^2R=0,
\end{eqnarray}
with
\begin{eqnarray}\label{II19}
\epsilon^2=\mathcal{E}^2-m^2-(1-c^2g)k^2.
\end{eqnarray}

We can note that the Eq. (\ref{II18a}) is analogous to the Eq. (\ref{II5}). Then, by following the same steps from the Eq. (\ref{II7}) to the Eq. (\ref{II10}), we obtain
\begin{eqnarray}\label{II20}
\mathcal{E}_{k,l,n}\approx\pm\sqrt{m^2+(1-c^2g)k^2+\frac{\pi^2}{\rho_0^2}\left(n+\frac{|l|}{2}+\frac{3}{4}\right)^2}.
\end{eqnarray}

The Eq. (17) represents the relativistic energy levels of a scalar field subject to a hard-wall potential in the spacetime with LSV governed by a background vector field in the $z$-direction. We can observe that the effects of the LSV influence the quantum dynamics of the scalar field through the presence of an effective linear momentum eigenvalue $k_{\text{eff}}=\sqrt{(1-c^2g)}k$. In addition, by making $g\rightarrow$ or $c\rightarrow0$, we obtain the relativistic energy levels of a scalar field subject to a hard-wall potential in the Minkowski spacetime.

\section{Effects of the Coulomb-type central potential}\label{SECIII}

The standard procedure of inserting central potentials into relativistic wave equations, such as the Klein-Gordon and Dirac equations, is through the minimum coupling which is represented by the transformation in the linear momentum operator, $\hat{p}_{\mu}\rightarrow\hat{p}_{\mu}-qA_{\mu}$, by $qA_0=V(\vec{r})=V(r)$. A well-known example, in a system of spherical symmetry, is the Coulomb potential in the description of the hydrogen atom and in the piônic atom \cite{greiner}. Another procedure to insert central potentials in relativistic wave equations, in particular in the Klein-Gordon equation, is by modifying the mass term of the equation, as shown in Ref. \cite{greiner}. Recently, through the modification of the mass term of the Klein-Gordon equation, some studies have been done in the context of quantum mechanics, for example, in exact solutions of the mass-dependent Klein-Gordon equation with the vector quark-antiquark interaction and harmonic oscillator potential \cite{by}, in the relativistic quantum dynamics of a charged particle in cosmic string spacetime in the presence of magnetic field and scalar potential \cite{eug}, in the linear confinement of a scalar particle in a G\"odel-type spacetime \cite{me5} and in exact solutions of the Klein–Gordon equation in the presence of a dyon, magnetic flux and scalar potential in the spacetime of gravitational defects \cite{eug1}. In this section, we take into account a scalar potential proportional to the inverse of the axial distance. It is important to mention that the Coulomb-type potential has been studied under the effects of the Klein-Gordon oscillator \cite{bf, me3}, in a Dirac particle \cite{pc}, with propagation of gravitational waves \cite{pc9}, in condensed matter systems, such as 1-dimensional systems \cite{pc1, pc2, pc3}, pseudo-harmonic interactions \cite{pc4, pc5} and molecules \cite{pc6, pc7, pc8}. Then, inspired by Ref. \cite{greiner}, we introduce a Coulomb-type central potential into the Klein-Gordon equation by modifying the mass term, $m\rightarrow m+U(\vec{r})$, where $m$ is a constant that corresponds to the rest mass of the scalar field and $U(\vec{r})$ is a scalar potential, with the intention of confining the scalar field in a spacetime with LSV and investigating the effects of the Coulomb-type central potential and spacetime anisotropies generated by the background vector field $v^{\mu}$ on the relativisitic quantum dynamics of a scalar field. Then, the mass term of the Klein-Gordon equation becomes $m\rightarrow m+\frac{\nu}{\rho}$, where $\nu$ is a constant that characterizes the Coulomb-type central potential. In this way, Eq. (\ref{II1}) takes the form
\begin{eqnarray}\label{III1}
\left[\square-g(v^{\mu}\partial_{\mu})^2-\left(m+\frac{\nu}{\rho}\right)^2\right]\phi=0.
\end{eqnarray}

\subsection{Background vector field with the configuration $v^{\mu}=(v^0,0)$}\label{SUBSECIII-1}

Let us consider a background vector field with the following configuration: $v^{\mu}=(a,0,0,0)$. In this particular case, the Eq. (\ref{III1}) becomes
\begin{eqnarray}\label{III2}
-\frac{\partial^2\phi}{\partial t^2}+\frac{\partial^2\phi}{\partial\rho^2}+\frac{1}{\rho}\frac{\partial\phi}{\partial\rho}+\frac{1}{\rho^2}\frac{\partial\phi}{\partial\varphi}+
\frac{\partial^2\phi}{\partial z^2}-ga^2\frac{\partial^2\phi}{\partial t^2}-\frac{2m\nu}{\rho}\phi-\frac{\nu^2}{\rho^2}\phi-m^2\phi=0.
\end{eqnarray}\label{III3}
By following the steps from the Eq. (\ref{II3}) to the Eq. (\ref{II5}), we obtain the axial wave equation
\begin{eqnarray}\label{III4}
\frac{d^2R}{d\rho^2}+\frac{1}{\rho}\frac{dR}{d\rho}-\frac{\gamma^2}{\rho^2}R-\frac{2m\nu}{\rho}R-\bar{\alpha}^2R=0,
\end{eqnarray}
where
\begin{eqnarray}\label{III5}
\bar{\alpha}^2=m^2+k^2-(1+a^2g)\mathcal{E}^2; \ \ \ \gamma^2=l^2+\nu^2.
\end{eqnarray}

Let us define $r=2\bar{\alpha}\rho$ , then the Eq. (\ref{III5}) becomes
\begin{eqnarray}\label{III6}
\frac{d^2R}{dr^2}+\frac{1}{r}\frac{dR}{dr}-\frac{\gamma^2}{r^2}R+\frac{\delta}{r}R-\frac{1}{4}R=0,
\end{eqnarray}
where
\begin{eqnarray}\label{III7}
\delta=\frac{m|\nu|}{\alpha}.
\end{eqnarray}
Next, let us impose that $R(r)\rightarrow0$ when $r\rightarrow0$ and $r\rightarrow\infty$. In this way, the radial wave function can be written as
\begin{eqnarray}\label{III8}
R(r)=r^{|\gamma|}e^{-\frac{1}{2}r}F(r),
\end{eqnarray}
then, we obtain the following equation for $F(r)$:
\begin{eqnarray}\label{III9}
r\frac{d^2F}{dr^2}+(2|\gamma|+1-r)\frac{dF}{dr}+\left(\delta-\frac{1}{2}-|\gamma|\right)F=0,
\end{eqnarray}
which is called in literature as the confluent hypergeometric equation and $F(r)=_{1}F_{1}\left(|\gamma|+\frac{1}{2}-\delta,2|\gamma|+1;r\right)$ is the confluent hypergeometric function \cite{arf}. It is well-known that the confluent hypergeometric series becomes a polynomial of degree $n$ by imposing that $|\gamma|+\frac{1}{2}-\delta=-n$, where $n=0,1,2,3,\ldots$. With this condition, we obtain
\begin{eqnarray}\label{III10}
\mathcal{E}_{k,l,n}=\pm\sqrt{\frac{1}{(1+a^2g)}\left[m^2+k^2-\frac{m^2\nu^2}{\left(n+\sqrt{l^2+\nu^2}+\frac{1}{2}\right)^2}\right]}.
\end{eqnarray}

Hence, by introducing the scalar potential by modification of the mass term, we can note, through Eq. (\ref{III10}) which represents the relativistic energy levels of the scalar field, the spectrum of energy is modified by the influence of the Coulomb-type central potential. Note also that the spacetime with LSV influences energy levels through the presence of parameters $a$ and $g$. By making $a\rightarrow0$ or $g\rightarrow0$, we have the relativistic energy levels of the scalar field subject to the Coulomb-type central potential in the Minkowski spacetime.

\subsection{Background vector field with the configuration $v^{\mu}=(0,\vec{v})$}\label{SUBSECIII-2}

\subsubsection{The axial direction}

Let us consider a external vector field with the following configuration: $v^{\mu}=(0,b,0,0)$. In this particular case, the axial wave equation becomes
\begin{eqnarray}\label{III11}
(1-b^2g)\frac{d^2R}{d\rho^2}+\frac{1}{\rho}\frac{dR}{d\rho}-\frac{\gamma^2}{\rho^2}R-\frac{2m\nu}{\rho}R+\beta^2R=0,
\end{eqnarray}
where $\beta^2$ and $\gamma^2$ are defined in the Eqs. (\ref{II13}) and (\ref{III5}), respectively. By substituting the Eq. (\ref{II14}) into the (\ref{III11}), we obtain
\begin{eqnarray}\label{III12}
\frac{d^2f}{d\rho^2}+\frac{1}{\rho}\frac{df}{d\rho}-\frac{\iota^2}{\rho^2}f-\frac{2m\nu}{(1-b^2g)\rho}f-\bar{\varepsilon}^2f=0,
\end{eqnarray}
where
\begin{eqnarray}\label{III13}
\iota^2=\frac{1}{4(1-b^2g)^2}(4\gamma^2-4\gamma^2b^2g+b^4g^2); \ \ \ \bar{\varepsilon}=\frac{\beta^2}{b^2g-1}.
\end{eqnarray}
The Eq. (\ref{III12}) is analogous to the Eq. (\ref{III6}). Then, by following the same steps from the Eq. (\ref{III8}) to the Eq. (\ref{III10}), we have
\begin{eqnarray}\label{III14}
\mathcal{E}_{k,l,n}=\pm\sqrt{m^2+k^2-(1-b^2g)\frac{m^2\nu^2}{\left(n+|\iota|+\frac{1}{2}\right)^2}},
\end{eqnarray}
which is the energy spectrum of a scalar field subject to the Coulomb-type central potential in a spacetime with LSV generated by a external external vector field in the axial direction. Note that the nature of the external vector field influences on energy levels through the presence of the parameters associated with the LSV, $b$ and $g$. By making $b\rightarrow0$  or $g\rightarrow0$, we obtain the relativistic energy levels of a scalar field subject to the Coulomb-type central potential in the Minkowski spacetime.

\subsubsection{z-direction}

Let us consider a background vector field with the following configuration: $v^{\mu}=(0,0,0,c)$. In this particular case, the axial wave equation becomes
\begin{eqnarray}\label{III15}
\frac{d^2R}{d\rho^2}+\frac{1}{\rho}\frac{dR}{d\rho}-\frac{\gamma^2}{\rho^2}R-\frac{2m\nu}{\rho}R-\bar{\epsilon}^2R=0,
\end{eqnarray}
where $\gamma^2$ is defined in the Eq. (\ref{III5}) and
\begin{eqnarray}\label{III16}
\bar{\epsilon}^2=m^2+(1-c^2g)k^2-\mathcal{E}^2.
\end{eqnarray}

We can note that the Eq. (\ref{III15}) is analogous to the Eq. (\ref{III4}). Then, by following the same steps from the Eq. (\ref{III4}) to the Eq. (\ref{III10}), we obtain
\begin{eqnarray}\label{III17}
\mathcal{E}_{k,l,n}=\pm\sqrt{m^2+(1-c^2g)k^2-\frac{m^2\nu^2}{\left(n+\sqrt{l^2+\nu^2}+\frac{1}{2}\right)^2}},
\end{eqnarray}
which is the general expression for the relativistic energy levels for the scalar field with position-dependent mass described in the Eq. (\ref{III1}) in the spacetime with LSV. We can observe the influence of the LSV in the Eq. (\ref{III17}) through the presence of the parameters $c$ and $g$. They yield a shift in the linear momentum eigenvalue that gives rise to an effective linear momentum quantum number $k_{\text{eff}}=\sqrt{1-c^2g}k$. In addition, by making $c\rightarrow0$ and $g\rightarrow0$, we obtain the relativistic energy levels of a scalar field subject to the Coulomb-type central potential in the Minkowski spacetime.

\section{Effects of the linear central potential}\label{SECIV}

In this section, we analyse the relativistic quantum effects of a linear central potential, through the modification of the mass term \cite{greiner} as $m\rightarrow m+\mu\rho$ \cite{me, eug}, where $\mu$ is a constant, and the effects of the LSV on the scalar field. The linear central potential has been studied investigated under the effects of the Klein-Gordon oscillator \cite{me2}, in the relativistic quantum dynamics of a scalar particle in the spacetime with torsion \cite{me4} and in Majorana fermion \cite{pl}. In this way, the a Klein-Gordon equation (\ref{II1}) becomes
\begin{eqnarray}\label{IV1}
[\square-g(v^{\mu}\partial_{\mu})^2-(m+\mu\rho)^2]\phi=0.
\end{eqnarray}

\subsection{Background vector field with the configuration $v^{\mu}=(v^0,0)$}\label{SUBSECIV-1}

Let us consider a background vector field with the following configuration: $v^{\mu}=(a,0,0,0)$. In this particular case, the axial wave equation becomes
\begin{eqnarray}\label{IV2}
\frac{d^2R}{d\rho^2}+\frac{1}{\rho}\frac{dR}{d\rho}-\frac{l^2}{\rho^2}R-2m\mu\rho R-\mu^2\rho^2R+\alpha^2R=0,
\end{eqnarray}
where $\alpha^2$ is defined in the Eq. (\ref{II6}).

From now on, let us consider $\xi=\sqrt{\mu}\rho$, thus, we rewrite the Eq. (\ref{IV2}) in the form
\begin{eqnarray}\label{IV3}
\frac{d^2R}{d\xi^2}+\frac{1}{\xi}\frac{dR}{d\xi}-\frac{l^2}{\xi^2}R-\eta\xi R-\xi^2R+\frac{\alpha^2}{\eta}R=0,
\end{eqnarray}
where we define the new parameter
\begin{eqnarray}\label{IV4}
\eta=\frac{2m}{\sqrt{\mu}}.
\end{eqnarray}
By analysing the asymptotic behaviour at $\xi\rightarrow0$ and $\xi\rightarrow\infty$, then, we can write the function $R(\xi)$ in terms of an unknown function $H(\xi)$ in the form \cite{me, me1, eug}:
\begin{eqnarray}\label{IV5}
R(\xi)=\xi^{l}e^{-\frac{1}{2}\xi(\xi+\eta)}H(\xi),
\end{eqnarray}
and thus, by substituting the Eq. (\ref{IV5}) into the Eq. (\ref{IV4}), we can observe that the function $H(\xi)$ is a solution to the following second order differential equation:
\begin{eqnarray}\label{IV6}
\frac{d^2H}{d\xi^2}+\left[\frac{(2|l|+1)}{\xi}-\eta-2\xi\right]\frac{dH}{d\xi}+\left[h-\frac{g}{\xi}\right]H=0,
\end{eqnarray}
where
\begin{eqnarray}\label{IV7}
h=\frac{\alpha^2}{\mu}-2-2|l|+\frac{\eta^2}{4}; \ \ \ g=\frac{\eta}{2}(2|l|+1).
\end{eqnarray}
The Eq. (\ref{IV7}) is called in the literature as the biconfluent Heun equation \cite{eug, heun} and the function $H(\xi)=H_{B}\left(2|l|,\eta,\frac{\alpha^2}{\mu}+\frac{\eta^2}{4},0;\xi\right)$ is the biconfluent Heun function.

Let us search for polynomial solutions to the Eq. (\ref{IV6}), then, for this purpose, we write the solution to the Eq. (\ref{IV6}) as a power series expansion around the origin, which is a regular singular point \cite{eug}:
\begin{eqnarray}\label{IV8}
H(\xi)=\sum_{j=0}^{\infty}d_j\xi^j.
\end{eqnarray}
By substituting this series into the Eq. (\ref{IV6}), we obtain the recurrence relation
\begin{eqnarray}\label{IV9}
d_{j+2}=\frac{[g+\eta(j+1)]d_{j+1}-(h-2j)d_j}{(j+2)(j+2+2|l|)},
\end{eqnarray}
where the coefficients $d_{1}$ and $d_{2}$ are
\begin{eqnarray}\label{IV10}
d_1&=&\frac{g}{(1+2|l|)}d_0=\frac{\eta}{2}; \\ \nonumber
d_2&=&\frac{(g+\eta)d_1-hd_0}{2(2+2|l|)}=\frac{1}{4(1+|l|)}\left[\frac{\eta^2}{4}(2|l|+3)-h\right],
\end{eqnarray}
with $d_0=1$.

In search of polynomial solutions to the biconfluent Heun equation (\ref{IV6}), we can note from the Eq. (\ref{IV8}) that the biconfluent Heun series becomes a polynomial of degree $\bar{n}$ when \cite{eug}
\begin{eqnarray}\label{IV11}
h=2\bar{n}; \ \ \ d_{\bar{n}+1},
\end{eqnarray}
where $\bar{n}=1,2,3,4,\ldots$. Therefore, the condition $h=2\bar{n}$ gives the expression
\begin{eqnarray}\label{IV12}
\mathcal{E}_{k,l,\bar{n}}=\pm\sqrt{\frac{1}{(1+a^2g)}[k^2+2\mu(1+\bar{n}+|l|)]}.
\end{eqnarray}

However, our analysis is not complete, since condition $d_{\bar{n}+1}=0$ must be analyzed and this can only be attributed by values of $\bar{n}$ in it. In this case, considering $\bar{n}=1$, which from the physical point of view represents the lowest energy state of the system and choosing the parameter associated to the linear potential $\mu=\mu_{k,l,\bar{n}}$ to adjust condition $d_{\bar{n}+1}=0$, that is, $d_{\bar{n}+1}=d_2=0$, not only for $\bar{n}=1$ but for any value of $\bar{n}$, we obtain the allowed values from $\mu$ to the radial mode $\bar{n}=1$:
\begin{eqnarray}\label{IV13}
\mu_{l,1}=\frac{m^2}{2}(2|l|+3).
\end{eqnarray}
With the relation given in the Eq. (\ref{IV13}), we have that the possible values of the parameter $\mu$ are determined by the quantum numbers $\{l,\bar{n}\}$ of the system. By substituting the Eq. (\ref{IV13}) into the Eq. (\ref{IV12}), the allowed energies for the lowest energy state are given by
\begin{eqnarray}\label{IV14}
\mathcal{E}_{k,l,1}=\pm\sqrt{\frac{1}{(1+a^2g)}[m^2(3+2|l|)(2+|l|)+k^2]}.
\end{eqnarray}
It is important to note that, unlike the previous cases analyzed, it is not possible to determine a closed solution for the biconfluent Heun polynomials for its more general case of its asymptotic behavior, that is, for large values of its argument. Hence, through the two conditions given in the Eq. (\ref{IV11}), arising from truncation of the power series (\ref{IV8}), it is only possible to determine polynomial solutions separately for each radial mode $\bar{n}$ of the system, as discussed in the Refs. \cite{eug, heun, heun1, heun2}. From the physical point of view, this quantum effect arises due to the presence of the linear central potential in the system. Besides, we can note the influence of the LSV in the Eq. (\ref{IV14}) through the presence of the parameters $a$ and $g$. By taking $a\rightarrow0$ or $g\rightarrow0$ in the Eq. (\ref{IV14}), we obtain  the allowed energies for the lowest energy state for the position-dependent mass system in the Minkowski spacetime.

\subsection{Background vector field with the configuration $v^{\mu}=(0,\vec{v})$}\label{SUBSECIV-2}

\subsubsection{The axial direction}

Let us consider a external vector field with the following configuration: $v^{\mu}=(0,b,0,0)$. In this particular case, the axial wave equation becomes
\begin{eqnarray}\label{IV15}
(1-b^2g)\frac{d^2R}{d\rho^2}+\frac{1}{\rho}\frac{dR}{d\rho}-\frac{l^2}{\rho^2}R-2m\mu\rho R-\mu^2\rho^2R+\beta^2R=0,
\end{eqnarray}
where $\beta^2$ is defined in the Eq. (\ref{II13}). By substituting the Eq. (\ref{II14}) into the Eq. (\ref{IV15}), we obtain
\begin{eqnarray}\label{IV16}
\frac{d^2f}{d\rho^2}+\frac{1}{\rho}\frac{df}{d\rho}-\frac{\lambda^2}{\rho^2}f-\frac{2m\mu\rho}{(1-b^2g)}f-\frac{\mu^2\rho^2}{(1-b^2g)}f+\frac{\beta^2}{(1-b^2g)}f=0,
\end{eqnarray}
where $\lambda^2$ is defined in the Eq. (\ref{II16}).

Let us define $s=\frac{\sqrt{\mu}}{(1-b^2g)^{1/4}}\rho$, then the Eq. (\ref{IV16}) becomes
\begin{eqnarray}\label{IV17}
\frac{d^2f}{ds^2}+\frac{1}{s}\frac{df}{ds}-\frac{\lambda^2}{s^2}-\bar{\eta}sf-s^2f+\Lambda f=0,
\end{eqnarray}
where
\begin{eqnarray}\label{IV18}
\bar{\eta}=\frac{2m}{\sqrt{\mu}(1-b^2g)^{1/4}}; \ \ \ \Lambda=\frac{\beta^2}{\mu\sqrt{1-b^2g}}.
\end{eqnarray}

We can note that the Eq. (\ref{IV18}) is analogous to the Eq. (\ref{IV3}). Then, by following the steps from the Eq. (\ref{IV5}) to the Eq. (\ref{IV9}), we obtain the recurrence relation:
\begin{eqnarray}\label{IV19}
d_{j+2}=\frac{[\bar{g}+\bar{\eta}(j+1)]d_{j+1}-(\bar{h}-2j)d_j}{(j+2)(j+2+2|l|)},
\end{eqnarray}
with the relations
\begin{eqnarray}\label{IV20}
d_1&=&\frac{\bar{g}}{(1+2|\lambda|)}=\frac{\bar{\eta}}{2}; \nonumber \\
d_2&=&\frac{1}{4(1+|\lambda|)}\left[\frac{\bar{\eta}^2}{4}(2|\lambda|+1)+\frac{\bar{\eta}^2}{2}-\bar{h}\right],
\end{eqnarray}
where we are considering $d_0=1$ and define the new parameters
\begin{eqnarray}\label{IV21}
\bar{h}=\Lambda-2-2|\lambda|+\frac{\bar{\eta}^2}{4}; \ \ \ \bar{g}=\frac{\bar{\eta}}{2}(2|\lambda|+1).
\end{eqnarray}

In search of a polynomial solution to the function $f(s)$, we have that the polynomial of degree $\bar{n}$ to $f(s)$ is achieved when we impose that \cite{eug}
\begin{eqnarray}\label{IV22}
\bar{h}=2\bar{n}; \ \ \ d_{\bar{n}+1}=0,
\end{eqnarray}
where $\bar{n}=1,2,3,4,\ldots$. From the condition $\bar{h}=2\bar{n}$, we have the expression
\begin{eqnarray}\label{IV23}
\mathcal{E}_{k,l,\bar{n}}=\pm\sqrt{k^2+2\sqrt{1-b^2g}\mu_{k,l,\bar{n}}(1+\bar{n}+|\lambda|)},
\end{eqnarray}
where we have labelled $\mu=\mu_{k,l,\bar{n}}$ as in the previous section. Further, let us analyse the condition $d_{\bar{n}+1}=0$ by working with the
lowest energy state $\bar{n}=1$. In this case, we have that $d_{\bar{n}+1}=d_2=0$, and then, the possible values of the parameter $\mu$ are determined by
\begin{eqnarray}\label{IV24}
\mu_{l,1}=\frac{m^2(2|\lambda|+3)}{2\sqrt{1-b^2g}}.
\end{eqnarray}

Note that the allowed values of $\mu$ are determined by the quantum numbers of the system $\{l,\bar{n}\}$ and parameters associated with the LSV, $b$ and $g$, in contrast with the previous subsection (\ref{IV13}). Hence, the allowed energies for the lowest energy state ($\bar{n}=1$) are
\begin{eqnarray}\label{IV25}
\mathcal{E}_{k,l,1}=\pm\sqrt{m^2\left[3+\frac{\sqrt{4l^2-4l^2b^2g+b^4g^2}}{(1-b^2g)}\right]\left[2+\frac{\sqrt{4l^2-4l^2b^2g+b^4g^2}}{2(1-b^2g)}\right]+k^2}.
\end{eqnarray}

The Eq. (\ref{IV25}) is the expression of the relativistic energy levels of the lowest energy state, $\bar{n}=1$, for a scalar field subject to a linear central potential in the spacetime with LSV caused by the presence of a external vector field. We can observe the influence of the LSV in the Eqs. (\ref{IV24}) and (\ref{IV25}) through the presence of the parameters $b$ and $g$. In addition, by making $b\rightarrow0$ or $g\rightarrow0$, we obtain the expression of the relativistic energy level of the lowest energy state in the Minkowski spacetime.

\subsubsection{z-direction}

Now, let us consider the configuration of the vector field given in the form $v^{\mu}=(0,0,0,c)$. In this particular case, the axial wave equation becomes
\begin{eqnarray}\label{IV26}
\frac{d^2R}{d\rho^2}+\frac{1}{\rho}\frac{dR}{d\rho}-\frac{l^2}{\rho^2}R-2m\mu\rho R-\mu^2\rho^2R+\epsilon^2R=0,
\end{eqnarray}
where $\epsilon^2$ is defined in the Eq. (\ref{II19}).

The Eq. (\ref{IV26}) is analogous to the Eq. (\ref{IV2}). Then, following the same steps from the Eq. (\ref{IV2}) to the Eq. (\ref{IV14}), we obtain
\begin{eqnarray}\label{IV27}
\mathcal{E}_{k,l,1}=\pm\sqrt{m^2(3+2|l|)(2+|l|)+(1-c^2g)k^2},
\end{eqnarray}
which is the general expression for the allowed energies for the lowest energy state for the position-dependent mass system described in the Eq. (\ref{IV1}) in the spacetime with LSV. We can note the influence of the spacetime with LSV in the Eq. (\ref{IV27}) through the presence of the parameters $c$ and $g$. They yield a shift in the linear momentum quantum number that gives rise to an effective linear momentum quantum number $k_{\text{eff}}=\sqrt{1-c^2g}k$. By taking $c\rightarrow0$ or $g\rightarrow0$ in the Eq. (\ref{IV27}), we obtain the relativistic allowed energy levels for $\bar{n}=1$ of a scalar field subject to the linear central potential in the Minkowski spacetime.

\section{Effects of the Coulomb-type plus linear central potential}\label{SECV}

In this section, let us consider the scalar field in $(3+1)$ dimensions given in the Eq. (\ref{II1}) under the influence of the linear and Coulomb-type potentials, where they are inserted into the Klein-Gordon equation by modifying the mass term, $m\rightarrow m+\mu\rho+\frac{\nu}{\rho}$, that is, a Cornell-type potential \cite{by}, since we are working with cylindrical symmetry. This type of potential has been studied in the Refs. \cite{eug, eug1, me2}. In this way, the Klein-Gordon equation (\ref{II1}) becomes
\begin{eqnarray}\label{V1}
\left[\Box-(v^{\mu}\partial_{\mu})^2-\left(m+\mu\rho+\frac{\nu}{\rho}\right)^2\right]\phi=0.
\end{eqnarray}

\subsection{Background vector field with the configuration $v^{\mu}=(v^0,0)$}\label{SUBSECV-1}

Let us consider the background vector field $v^{\mu}=(a,0,0,0)$. In this particular case, the Eq. (\ref{V1}) gives axial wave equation
\begin{eqnarray}\label{V2}
\frac{d^2R}{d\rho^2}+\frac{1}{\rho}\frac{dR}{d\rho}-\frac{\gamma^2}{\rho^2}R-\frac{2m\nu}{\rho}R-2m\mu\rho R-\mu^2\rho^2R+\tilde{\alpha}^2R=0,
\end{eqnarray}
where $\gamma^2$ is defined in the Eq. (\ref{III5}) and we define the new parameter
\begin{eqnarray}\label{V3}
\tilde{\alpha}^2=(1+a^2g)\mathcal{E}^2-m^2-k^2-2\mu\nu.
\end{eqnarray}

From now on, let us consider $\xi=\sqrt{\mu}\rho$, thus, we rewrite the Eq. (\ref{V3}) in the form
\begin{eqnarray}\label{V4}
\frac{d^2R}{d\xi^2}+\frac{1}{\xi}\frac{dR}{d\xi}-\frac{\gamma^2}{\xi^2}R-\frac{\zeta}{\xi}R-\eta\xi R-\xi^2R+\frac{\tilde{\alpha}^2}{\mu}R=0,
\end{eqnarray}
where $\eta$ is defined in the Eq. (\ref{IV4}) and
\begin{eqnarray}\label{V5}
\zeta=\frac{2m\nu}{\sqrt{\mu}}.
\end{eqnarray}

By analysing the asymptotic behaviour of the possible solutions to the Eq. (\ref{V4}) at $\xi\rightarrow0$ and $\xi\rightarrow\infty$, we can write the function $R(\xi)$ in terms of an unknown function $G(\xi)$ as
\begin{eqnarray}\label{V6}
R(\xi)=\xi^{|\gamma|}e^{-\frac{1}{2}\xi(\xi+\eta)}G(\xi).
\end{eqnarray}
Then, by substituting the Eq. (\ref{V6}) into the Eq. (\ref{V4}), we obtain the biconfluent Heun equation \cite{eug, heun}
\begin{eqnarray}\label{V7}
\frac{d^2G}{d\xi^2}+\left[\frac{(2|\gamma|+1)}{\xi}-\eta-2\xi\right]\frac{dG}{d\xi}+\left[\tilde{h}-\frac{\tilde{g}}{\xi}\right]G=0,
\end{eqnarray}
where
\begin{eqnarray}\label{V8}
\tilde{h}=\frac{\tilde{\alpha}^2}{\mu}+\frac{\eta^2}{4}-2-2|\gamma|; \ \ \ \tilde{g}=\frac{\eta}{2}(2|\gamma|+1)+\zeta,
\end{eqnarray}
and the function $G(\xi)$ is the biconfluent Heun function, $G(\xi)=G_{B}\left(2|\gamma|,\eta,\frac{\tilde{\alpha}^2}{\mu}+\frac{\eta^2}{4},2\zeta;\xi\right)$.

Further, by using the Fr\"obenius method as in the Eqs. (\ref{IV8}) and (\ref{IV9}), we obtain the recurrence relation:
\begin{eqnarray}\label{V9}
d_{j+2}=\frac{[\tilde{g}+\eta(j+1)]d_{j+1}-(\tilde{h}-2j)d_j}{(j+2)(j+2+2|\gamma|)},
\end{eqnarray}
with the relations
\begin{eqnarray}\label{V10}
d_1&=&\frac{\tilde{g}}{1+2|\gamma|}; \\ \nonumber
d_2&=&\frac{1}{4(1+|\gamma|)}[(\tilde{g}+\eta)d_1-\tilde{h}]=\frac{1}{4(1+|\gamma|)}\left[\frac{\tilde{g}^2+\eta\tilde{g}}{1+2|\gamma|}-\tilde{h}\right],
\end{eqnarray}
where we are considering $d_0=1$.

As seen in the Sec. (\ref{SECIV}), through the recurrence relation (\ref{V9}), we can see that the power series expansion (\ref{V9}) becomes a polynomial of degree $\bar{n}$ by imposing two conditions \cite{eug}
\begin{eqnarray}\label{V11}
\tilde{h}=2\bar{n}; \ \ \ d_{\bar{n}+1}=0,
\end{eqnarray}
where $\bar{n}=1,2,3,4,\ldots$. From the condition $\tilde{h}=2\bar{n}$, we obtain the expression
\begin{eqnarray}\label{V12}
\mathcal{E}_{k,l,\bar{n}}=\pm\sqrt{\frac{1}{(1+a^2g)}[k^2+2\mu_{k,l,\bar{n}}(1+\bar{n}+|\gamma|+\nu)]}.
\end{eqnarray}
For our analysis to be complete, it is necessary to analyze the condition $d_{\bar{n}+1}=0$, which, as already seen in the Sec. (\ref{SECIV}), the parameter associated with the linear central potential is chosen to adjust such condition $(\mu=\mu_{k,l,\bar{n}})$. Then, for $\bar{n}=1$, we have $d_{\bar{n}+1}=d_2=0$, which it gives the allowed values of $\mu$ for $\bar{n}=1$
\begin{eqnarray}\label{V13}
\mu_{l,1}=\frac{m^2}{2}(2\sqrt{l^2+\nu^2}+3)+4m^2\nu\frac{(\sqrt{l^2+\nu^2}+1)}{(2\sqrt{l^2+\nu^2}+1)}+\frac{2m^2\nu^2}{(2\sqrt{l^2+\nu^2}+1)}.
\end{eqnarray}

By comparing the expressions for the allowed values of $\mu$ (\ref{IV13}) and (\ref{V13}), we have that the presence of the Coulomb-type potential modifies the the expression for the allowed values of $\mu$ for $\bar{n}=1$, that is, the possible values of $\mu$ are determined by the parameter associated with the Coulomb-type potential, and by the the quantum numbers of the system $\{l,n\}$. With the result given in the Eq. (\ref{V13}), the allowed energies for the lowest energy state are
\begin{eqnarray}\label{V14}
\mathcal{E}_{k,l,1}=\pm\sqrt{\frac{1}{(1+a^2g)}\left\{k^2+m^2\left[(2|\gamma|+3)+8\nu\frac{(|\gamma|+1)}{(2|\gamma|+1)}+\frac{4\nu^2}{(2|\gamma|+1)}\right](2+|\gamma|+\nu)\right\}}.
\end{eqnarray}

We can note that the allowed energy values for $\bar{n}=1$ are influenced by the spacetime with LSV through the presence of the constant parameters $a$ and $g$. Note that, by making $\nu=0$ in the Eqs. (\ref{V13}) and (\ref{V14}) we recover the Eqs. (\ref{IV13}) and (\ref{IV14}), respectively. In addition, by making $a\rightarrow0$ or $g\rightarrow0$, we obtain the allowed energy values of a scalar field subject to the Coulomb-type plus linear central potential in the Minkowski spacetime.

\subsection{Background vector field with the configuration $v^{\mu}=(0,\vec{v})$}\label{SUBSECV-2}

\subsubsection{The axial direction}

Let us consider the external vector field $v^{\mu}=(0,b,0,0)$. In this particular case, the Eq. (\ref{V1}) gives axial wave equation
\begin{eqnarray}\label{V15}
(1-b^2g)\frac{d^2R}{d\rho^2}+\frac{1}{\rho}\frac{dR}{d\rho}-\frac{\gamma^2}{\rho^2}R-\frac{2m\nu}{\rho}R-2m\mu\rho R-\mu^2\rho^2R+\bar{\beta}^2R=0
\end{eqnarray}
where $\gamma^2$ is defined in the Eq. (\ref{III5}) and
\begin{eqnarray}\label{V16}
\bar{\beta}^2=\mathcal{E}^2-m^2-k^2-2\mu\nu.
\end{eqnarray}
By substituting the Eq. (\ref{II14}) into the Eq. (\ref{V15}), we obtain
\begin{eqnarray}\label{V17}
\frac{d^2f}{d\rho^2}+\frac{1}{\rho}\frac{df}{d\rho}-\frac{\iota^2}{\rho^2}f-\frac{2m\nu}{(1-b^2g)\rho}f-\frac{2m\mu}{(1-b^2g)}\rho f-\frac{\mu^2}{(1-b^2g)}f+\frac{\bar{\beta}^2}{(1-b^2g)}f=0,
\end{eqnarray}
where $\iota^2$ is defined in the Eq. (\ref{III13}). Now, by using the variable change $s=\frac{\sqrt{\mu}}{(1-b^2g)^{1/4}}\rho$, we have
\begin{eqnarray}\label{V18}
\frac{d^2f}{ds^2}+\frac{1}{s}\frac{df}{ds}-\frac{\iota^2}{s^2}f-\frac{\bar{\zeta}}{s}f-\bar{\eta}sf-s^2f+\bar{\Lambda}f=0,
\end{eqnarray}
where $\bar{\eta}$ is defined in the Eq. (\ref{IV18}) and
\begin{eqnarray}\label{V19}
\bar{\zeta}=\frac{2m\nu}{\sqrt{\mu}(1-b^2g)^{1/4}}; \ \ \ \bar{\Lambda}=\frac{\bar{\beta}^2}{\mu(1-b^2g)}.
\end{eqnarray}

We can note that the Eq. (\ref{V18}) is analogous to the Eq. (\ref{V4}). Then, let us follow the steps from the Eq. (\ref{V6}) to the Eq. (\ref{V12}), we obtain the expression
\begin{eqnarray}\label{V20}
\mathcal{E}_{k,l,\bar{n}}=\pm\sqrt{k^2+2\sqrt{1-b^2g}\mu_{k,l,\bar{n}}(1+\bar{n}+|\iota|+\nu)}.
\end{eqnarray}

As we have discussed in the previous section, the parameter $\mu$ can be adjusted to satisfy the truncation conditions of the biconfluent Heun series. Then, for $\bar{n}=1$, we obtain the allowed values of $\mu$
\begin{eqnarray}\label{V21}
\mu_{l,1}=\frac{m^2(2|\iota|+3)}{2\sqrt{1-b^2g}}+\frac{4m^2\nu(|\iota|+1)}{(1-b^2g)(1+2|\iota|)}+\frac{2m^2\nu^2}{(1-b^2g)^{3/2}(1+2|\iota|)}.
\end{eqnarray}

We can observe that, in contrast to the Subsec. (\ref{SUBSECV-1}), the configuration of the external vector field that characterizes the LSV modifies the allowed values of $\mu$ for $n=1$. By making $\nu=0$, we recover the Eq. (\ref{IV24}). Hence, the allowed values of the relativistic energy for the radial mode $\bar{n}=1$ are
\begin{eqnarray}\label{V22}
\mathcal{E}_{k,l,1}=\pm\sqrt{k^2+m^2\left[(2|\iota|+3)+\frac{8\nu(1+|\iota|)}{\sqrt{1-b^2g}(1+2|\iota|)}+\frac{4\nu^2}{(1-b^2g)(1+2|\iota|)}\right](2+|\iota|+\nu)}.
\end{eqnarray}

We can note the influence of the LSV in the Eq. (\ref{V22}) through the presence of the parameters $b$ and $g$. In addition, by making $\nu=0$, we recover Eq. (\ref{IV25}). By talking $b\rightarrow0$ or $g\rightarrow0$, we obtain the allowed energy for $\bar{n}=1$ of a scalar field subject to the Coulomb-type plus linear central potential in the Minkowski spacetime.

\subsubsection{z-direction}

Let us consider the background vector field $v^{\mu}=(0,0,0,c)$. In this particular case, the Eq. (\ref{V1}) gives axial wave equation becomes
\begin{eqnarray}\label{V23}
\frac{d^2R}{d\rho^2}+\frac{1}{\rho}\frac{dR}{d\rho}-\frac{\gamma^2}{\rho^2}R-\frac{2m\nu}{\rho}R-2m\mu\rho R-\mu^2\rho^2R+\tilde{\epsilon}^2R=0
\end{eqnarray}
where $\gamma^2$ is defined in the Eq. (\ref{III5}) and
\begin{eqnarray}\label{V24}
\tilde{\alpha}^2=\mathcal{E}^2-m^2-(1-c^2g)k^2-2\mu\nu.
\end{eqnarray}

The Eq. (\ref{V23}) is analogous to the Eq. (\ref{V2}). Then, by following the same steps from the Eq. (\ref{V2}) to the Eq. (\ref{V14}), we obtain
\begin{eqnarray}\label{V25}
\mathcal{E}_{k,l,1}=\pm\sqrt{(1-c^2g)k^2+m^2\left[(2|\gamma|+3)+\frac{8\nu(|\gamma|+1)}{(2|\gamma|+1)}+\frac{4\nu^2}{(2|\gamma|+1)}\right](2+|\gamma|+\nu)},
\end{eqnarray}
which is the general expression for the allowed energies for the lowest energy state of the system described in the Eq. (\ref{V1}) in the spacetime with LSV. We can note the influence of the LSV in the Eq. (\ref{V25}) through the presence of the parameters $c$ and $g$. They yield a shift in the linear momentum eigenvalue that gives rise to an effective linear momentum quantum number $k_{\text{eff}}=\sqrt{1-c^2g}k$. In addition, by making $\nu=0$ in the Eq. (\ref{V25}), we recover the Eq. (\ref{IV27}). By taking $c\rightarrow0$ or $g\rightarrow0$ in the Eq. (\ref{V25}), we obtain  the allowed energy values for $\bar{n}=1$ of a scalar field subject to the Coulomb-type plus linear central potential in the Minkowski spacetime.

\section{Conclusions}

We have investigated the effects of the LSV on a scalar field subject to a hard-wall potential and Coulomb-type and linear central potentials. The LSV is governed by the presence of a background constant vector field which modifies the structure of the Klein-Gordon equation (\ref{II1}) by it is directly coupled to a derivative of the field. This coupling conserves the $CPT$ symmetry, so it is considered a $CPT$-even coupling \cite{go}. For our analysis, we consider the particular cases where the vector field has the particular configurations $v^{\mu}=(v^0,0)$ and $v^{\mu}=(0,\vec{v})$. In the particular case where the vector field has the particular configuration $v^{\mu}=(0,\vec{v})$, it is possible to note that, by a coordinate change, we obtain the Klein-Gordon equation in its ordinary form. However, we are determining explicit results of this background on the scalar field, since we are interested in a possible detection.

In our first analysis, we considered the presence of a hard-wall potential, where we have shown that there is the influence of the effects of the LSV on the relativistic energy levels. Then, by modifying the mass term, we inserted the Coulomb-type central potential into the Klein-Gordon equation, where we determined the energy levels of the analyzed systems, which in turn in all cases we can note the influence of the LSV on the levels of the relativistic energy levels. In the case of the linear central potential, we have calculated the values allowed for the lower energy states of the system and shown that there is also influence of the LSV. Then we extend our analysis considering the presence of the Coulom-type potential plus the linear potential and show that relativistic energy allowed for lower energy state is affected by the effects of the LSV. In addition, the influence of the linear central potential and of the Coulomb-type plus linear central potential on the scalar field restricts the values of the parameter related to the linear central potential to a set of values that are established by the quantum numbers of the system which allow us to obtain a polynomial solution to the biconfluent Heun series. We also can note that because the symmetry is cylindrical, the scalar field is subject to the effects of the axial central potentials only in the $xy$-plane.

It is worth mentioning that the background vector field introduced in the Klein-Gordon equation can be considered more general, that is, where all its components are non-zero. It is in our interest as future perspectives to analyze this more general case on the scalar field, not only for the central potentials considered in the present work, but for other interactions and external effects, for example, the Klein-Gordon oscillator \cite{okg, okg1, okg2, okg3, okg4}, the Landau quantization \cite{landau}, the Aharonov-Bohm effect for bound states \cite{ab} and thermodynamic properties \cite{pt, pt1, pt2, pt3, pt4, pt5}.

\acknowledgments{The authors would like to thank the Brazilian agencie CNPq for financial support. R. L. L. Vit\'oria was supported by the CNPq project No. 150538/2018-9.}

\end{document}